\documentclass[showpacs,preprintnumbers,twocolumn,prl]{revtex4-1}
\usepackage{amsmath,amssymb}
\usepackage[dvips]{graphicx}

\newcommand{\MeV}{\;\text{MeV}}
\newcommand{\Nc}{N_{\text{c}}}
\newcommand{\Nf}{N_{\text{f}}}
\newcommand{\muB}{\mu_{\text{B}}}
\newcommand{\muq}{\mu_{\text{q}}}
\newcommand{\sfree}{s_{\text{free}}}
\newcommand{\nfree}{n_{\text{free}}}
\newcommand{\U}{\mathcal{U}}
\newcommand{\bPhi}{\bar{\Phi}}

\begin{document}
\title{Phase diagram of hot and dense QCD constrained by the
  Statistical Model}
\author{Kenji Fukushima}
\affiliation{Yukawa Institute for Theoretical Physics,
 Kyoto University, Kyoto 606-8502, Japan}
\begin{abstract}
 We propose a prescription to constrain the chiral effective model
 approach to the QCD phase diagram using the thermal Statistical
 Model, which is a description consistent with the experimental data
 at the freeze-out.  In the transition region where thermal quantities
 of hadrons blow up, deconfined quarks and gluons should smoothly take
 over the degrees of freedom from hadrons in the Statistical Model.
 We use the Polyakov-loop coupled Nambu--Jona-Lasinio (PNJL) model as
 an effective description in the quark side.  We require that the
 validity regions of these descriptions should have an overlap on the
 phase diagram, which constrains model uncertainty.  Our results
 favor a phase diagram with the chiral phase transition located at
 slightly higher temperature than deconfinement.
\end{abstract}
\preprint{YITP-10-46}
\maketitle


\paragraph*{Introduction}
Exploration of the QCD (Quantum Chromodynamics) phase diagram,
particularly toward higher baryon-density regime, is of increasing
importance in both theoretical and experimental
sides~\cite{Fukushima:2010bq}.  From the theoretical point of view, so
far, only the lattice-QCD
simulation~\cite{Fukushima:2010bq,DeTar:2009ef} is the first-principle
approach at work to the QCD phase transitions --- chiral restoration
and quark deconfinement.  The functional renormalization group method
is also developing as a promising non-perturbative
tool~\cite{Braun:2007bx}.  The chiral condensate
$\langle\bar{\psi}\psi\rangle$ and the Polyakov loop $\Phi$ are the
(approximate) order parameters for chiral restoration and quark
deconfinement, respectively, which are gauge invariant and measurable
on the lattice.  The lattice-QCD simulation is, however, of practical
use only when the baryon chemical potential $\muB$ is sufficiently
smaller than the temperature $T$.  For $\muB/T\gtrsim 1$ the notorious
sign problem prevents us from extracting any reliable information from
the lattice-QCD data~\cite{Fukushima:2010bq,Muroya:2003qs}.

The effective model study is an alternative and pragmatic approach
toward the phase diagram of dense QCD.\ \ The idea is the following;
one starts with some models that yield a reasonable description of
hadron properties in the vacuum and then puts them in a finite-$T$
and/or finite-$\muB$ environment.  What is recognized nowadays as the
``QCD phase diagram'' is actually a theoretical conjecture based on
various effective model studies.

Along this line the Polyakov-loop coupled chiral models such as the
PNJL
(Polyakov--Nambu--Jona-Lasinio)~\cite{Fukushima:2003fw,Ratti:2005jh}
and the PQM
(Polyakov-Quark-Meson)~\cite{Schaefer:2007pw,Schaefer:2009ui} models
are successful to handle $\langle\bar{\psi}\psi\rangle$ and $\Phi$ on
the equal footing.  Besides, the Polyakov loop potential
$\mathcal{U}[\Phi]$ is determined by $\Phi$ and the pressure $p$
measured in the lattice simulation of the pure gluonic theory.  This
means that the model includes the pressure contribution from gluons as
well as quarks, so that the model is able to deal with the full
thermodynamics comparable with the full lattice-QCD simulation.  The
point is that the dynamics of transverse gluons $A_i^T$ is under the
control of the deconfinement order parameter $\Phi$ and thus is to be
encompassed in the Polyakov loop potential $\mathcal{U}[\Phi]$, while
the Polyakov loop itself is expressed in terms of the longitudinal
gluon $A_4$. 

Since theory instruments to examine both
$\langle\bar{\psi}\psi\rangle$ and $\Phi$ are now in our hands, it is
intriguing to address the following question;  whether the chiral and
deconfinement phase transitions would go on simultaneously or separate
after all when the baryon density increases.  There are then two key
issues.  One is the so-called QCD (chiral) \textit{critical point}
(which is often called the \textit{critical end-point}) at which the
chiral and the baryon number susceptibilities
diverge~\cite{Asakawa:1989bq,Stephanov:1998dy,Son:2004iv} and the
higher moments are even more singular~\cite{Stephanov:2008qz}.  The
other one is a \textit{triple-point-like region} associated with the
appearance of quarkyonic matter~\cite{McLerran:2007qj,Andronic:2009gj}
where the baryon abundance surpasses mesons.

One reasonable way to characterize quarkyonic matter for finite-$\Nc$
QCD is to use two order parameters $\Phi=0$ and the quark (baryon)
number density $\langle\psi^\dagger \psi\rangle\neq0$, which would
definitely work for $\Nc=\infty$~\cite{McLerran:2007qj}.  In principle
this statement is not directly related to chiral symmetry, but a
substantially large value of $\langle\psi^\dagger\psi\rangle$ is
favored by light quarks existing in the chiral symmetric phase.  In
this sense, practically, one can identify the quarkyonic phase as an
exotic state where chiral symmetry is restored first
($\langle\bar{\psi}\psi\rangle\neq0$) and still the confining property
remains ($\Phi\simeq0$).  In other words the bulk pressure is mostly
dominated by light quarks and, nevertheless, excited quarks on top of
the Fermi sphere feel a confining force.  [There is an argument that
  the confining force might   cause inhomogeneous chiral
  condensation~\cite{Kojo:2009ha}.  Such a possibility is beyond our
  current scope.]

There is no strong evidence for such an exotic window.
Phenomenological considerations could lead to a different
scenario~\cite{Castorina:2010gy}, though some suggested arguments have
been reported~\cite{Andronic:2009gj,Hands:2010gd} and some model
studies are supportive~\cite{Fukushima:2003fw,Sasaki:2006ww}.  In
general the PNJL and PQM models rather favor the quarkyonic picture;
the model predicts the deconfinement temperature weakly dependent on
$\muB$.  The Polyakov loop tends to be small for any $\muB$ as long as
$T$ is vanishingly small, whereas the chiral condensate melts at high
$\muB$.  However, the serious problem in any model studies is that the
model-parameter choice is largely uncertain.  The PNJL and PQM models
are not exceptions.  The situation is worse at higher $\muB$ because
the lattice-QCD data is unavailable then.  It should fatally depends
on model assumptions whether the phase diagram has the critical
point(s) and/or quarkyonic matter or even nothing at all.  To make any
solid statement, it is indispensable to impose some constraints on the
effective model.  In this work we attempt to deduce the phase
structure from the phenomenological point of view.
\vspace{3mm}


\begin{figure}
 \includegraphics[width=\columnwidth]{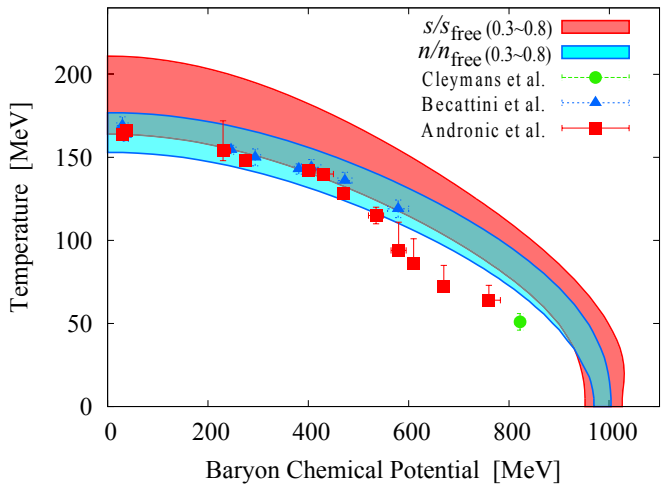}
 \caption{Chemical freeze-out points taken from
   Refs.~\cite{Cleymans:2005xv,Becattini:2005xt}.  The red and blue
   (upper and lower) bands represent the regions where the entropy
   density $s$ and the baryon number density $n$, respectively,
   increase quickly from $0.3$ to $0.8$ in the unit
   of free quark-gluon values, $\sfree$ and $\nfree$ (see
   Eq.~\eqref{eq:free}).}
 \label{fig:freezeout}
\end{figure}



\paragraph*{Thermodynamics from the Statistical Model}

Regarding the QCD phase diagram at finite $T$ and $\muB$ useful
information is quite limited.  Only the chemical freeze-out points in
the heavy-ion collisions are experimental hints about the phase
diagram.  Although the freeze-out points shape an intriguing curve on
the $\muB$-$T$ plane, as plotted by error-bar dots in
Fig.~\ref{fig:freezeout}, one should carefully interpret it.

The freeze-out points are not the raw experimental data but
\textit{an} interpretation through the Statistical
Model~\cite{Cleymans:2005xv,Becattini:2005xt}.  In view of the fact
that the Statistical Model is such successful to fit various particle
ratios with $\muB$ and $T$ only ($\mu_Q$, $\mu_{\text{s}}$, and
$\mu_{\text{c}}$ are determined by the initial condition), it should
be legitimate to take the freeze-out points for experimental
data, which in turn validates the Statistical Model (though why it
works lacks for an explanation from QCD).

Let us proceed further accepting that the Statistical Model is a valid
description of the state of matter until the freeze-out curve or
slightly above.  It is then a straightforward application of the
Statistical Model to estimate thermodynamic quantities such as the
pressure $p$, the entropy density $s$, the baryon number density $n$,
etc.  We shall utilize the open code THERMUS ver.2.1 to calculate $s$
and $n$ at various $T$ and $\muB$~\cite{Wheaton:2004qb}.

Figure~\ref{fig:freezeout} shows the chemical freeze-out points taken
from Refs.~\cite{Cleymans:2005xv,Becattini:2005xt}, on which $s$ and
$n$ are overlaid.  For convenience we normalized these quantities by
\begin{align}
 \sfree & = \biggl\{ (\Nc^2-1) + \frac{7}{4} \Nc\Nf\biggr\}
  \frac{4\pi^2}{45}T^3 + \frac{\Nc\Nf}{3} \muq^2 T \;,\notag\\
 \nfree &= \Nf \biggl( \frac{\muq^3}{3\pi^2}
  + \frac{\muq T^2}{3} \biggr) \;.
\label{eq:free}
\end{align}
These are the entropy density and the baryon number density of free
massless $\Nc^2-1$ gluons and $\Nc\Nf$ quarks.

Here we note that, in drawing Fig.~\ref{fig:freezeout}, we have
intentionally relaxed the neutrality conditions for electric charge
and heavy flavors and simply set
$\mu_Q=\mu_{\text{s}}=\mu_{\text{c}}=0$.  We have done so to make it
possible to compare the results from the Statistical Model to the
chiral effective model approach in later discussions.  [We note that
  one can force the chiral model to satisfy neutrality but it would be
  technically involved~\cite{Fukushima:2009dx}.]  Nevertheless, we
would emphasize that the neutrality conditions have only minor effects
on the bulk thermodynamics and make only small differences in any
case.

We should mention that we used Eq.~\eqref{eq:free} with $\Nc=\Nf=3$.
The choice of $\sfree$ and $\nfree$ is arbitrary and the following
discussions do not rely on this choice, for we will use $\sfree$ and
$\nfree$ just as common denominators to display the Statistical Model
and the PNJL model results.

The Statistical Model cannot tell us about the QCD phase transitions.
Still, Fig.~\ref{fig:freezeout} is suggestive enough.  We can clearly
see the thermodynamic quantities from the Statistical Model blowing up
in a relatively narrow region.  The red and blue (upper and lower)
bands indicate the regions where $s/\sfree$ and $n/\nfree$ ,
respectively, grow quickly from $0.3$ to $0.8$.  In the Hagedorn's
picture~\cite{Cabibbo:1975ig} this rapid and simultaneous rise in $s$
and $n$ has a natural interpretation as the Hagedorn limiting
temperature, above which color degrees of freedom is liberated,
i.e.\ color deconfinement.
\vspace{3mm}


\paragraph*{Thermodynamics from the PNJL Model}

Figure~\ref{fig:freezeout} is useful to have a guess-estimate about
the deconfinement boundary but we can deduce no information about the
chiral property.  So, to address the QCD phase transitions, we must
find a way to connect the thermodynamics in Fig.~\ref{fig:freezeout}
to the order parameters $\langle\bar{\psi}\psi\rangle$ and $\Phi$.
Here let us go into details of the chiral effective model for that
purpose.

It is essential to adopt the Polyakov-loop augmented model here
because the entropy density should contain the contribution from
gluons which is taken care of by the Polyakov loop potential.  The
PNJL model that we use in what follows is defined with the following
potential;
\begin{equation}
 \begin{split}
& \U[\Phi,\bPhi] = T^4 \biggl\{ -\frac{a(T)}{2}\bPhi\Phi \\
 &\quad + b(T)\ln\Bigl[ 1-6\bPhi\Phi+4(\bPhi^3+\Phi^3)
  -3(\bPhi\Phi)^2 \Bigr] \biggr\}
 \end{split}
\end{equation}
with $a(T)=a_0 + a_1(T_0/T) + a_2(T_0/T)^2$ and $b(T)=b_3(T_0/T)^3$.
There are thus five parameters one out of which is constrained by the
Stefan-Boltzmann limit.  These parameters are fixed by the
pure-gluonic lattice data as $a_0=3.51$, $a_1=-2.47$, $a_2=15.2$,
$b_3=-1.75$~\cite{Ratti:2005jh}, and $T_0=270\MeV$ from the
deconfinement temperature of first order in the pure-gluonic theory.
It is important to note that only $T_0$ is a parameter with the mass
dimension, so that the energy scale is set by this $T_0$.

In addition the NJL sector of the PNJL model has five more
parameters in the three-flavor case~\cite{Fukushima:2003fw}; the light
and heavy quark masses $m_{\rm ud}$ and $m_{\rm s}$, the momentum
cutoff $\Lambda$, the four-fermionic interaction strength $g_{\rm s}$,
and the $\mathrm{U(1)_A}$-breaking six-fermionic interaction strength
$g_{\rm d}$, which are all fixed by the pion mass $m_\pi$, the kaon
mass $m_K$, the eta-prime mass $m_{\eta'}$, the pion decay constant
$f_\pi$, and the chiral condensate
$\langle\bar{\psi}\psi\rangle$~\cite{Hatsuda:1994pi}.


\begin{figure}
 \includegraphics[width=\columnwidth]{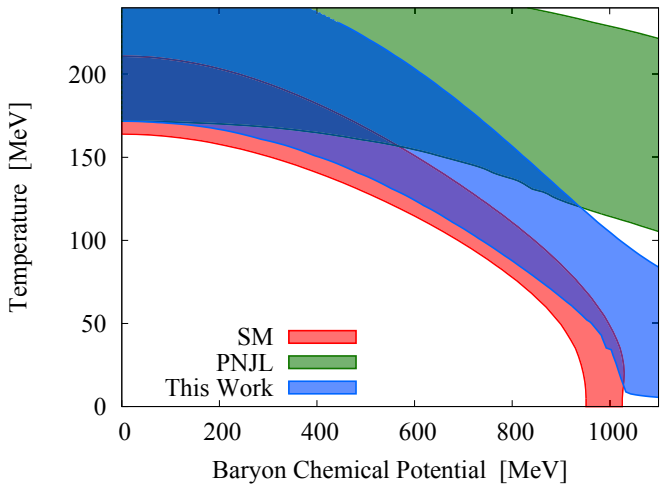}
 \caption{Entropy density normalized by $\sfree$ (from $0.3$ to $0.8$)
   in the Statistical Model (bottom band with red color; same as shown
   in Fig.~\ref{fig:freezeout}) and that in the PNJL model with a
   choice $T_0=200\MeV$ (top band with green color).  The blue band
   between two represents the results with the ansatz~\eqref{eq:T0}.}
 \label{fig:entropy}
\end{figure}


In the presence of dynamical quarks, if we keep using $T_0=270\MeV$,
the simultaneous crossover temperature of deconfinement and chiral
restoration is over $200\MeV$, which is too high as compared to the
lattice-QCD value.  It is argued in Ref.~\cite{Schaefer:2007pw} that
the back-reaction from quark loops affects the mass scale $T_0$ which
changes from $T_0=270\MeV$ for $\Nf=0$ reduced down to $T_0=208\MeV$
for $\Nf=2$ and $T_0=187\MeV$ for $\Nf=2+1$~\cite{Schaefer:2007pw}.
In this work we choose to use $T_0=200\MeV$ throughout.

In Fig.~\ref{fig:entropy} we show the entropy density calculated in
the mean-field PNJL model with $T_0=200\MeV$ in the same way as
presented in Fig.~\ref{fig:freezeout}.  The bottom (top) band in red
(green) color is the result from the Statistical Model (PNJL model).
From the figure it is obvious that the PNJL model is not consistent
with the Statistical Model even qualitatively.  With the properly
scaled $T_0$ from $270\MeV$ down to $200\MeV$, the blow-up behavior in
$s$ from the Statistical Model could be smoothly connected to the PNJL
model description only for $\muB\lesssim 400\MeV$.  The curvature
of the band as a function of $\muB$ is so different;  the PNJL model
result is too flat horizontally and it eventually take apart from the
region where the Statistical Model breaks down.
\vspace{3mm}


\paragraph*{Problem and Solution}

Such a manifest discrepancy from the Statistical Model is a crucial
drawback of the PNJL model.  To make the entropy density at
$\muB\gtrsim 500\MeV$ get saturated, quark degrees of freedom must be
released at smaller temperature than predicted by the PNJL model.

One can imagine why this happens in the following way;  the energy
scale in the Polyakov loop potential is specified by the parameter
$T_0$ that may differ depending on the effects of $T$ and $\muB$ in
the quark sector.  We have shifted $T_0$ from $270\MeV$ down to
$200\MeV$, through which we have incorporated the scale change induced
by $\Nf$ quarks at finite $T$.  In this way we may well consider that
$T_0$ should decrease with increasing
$\muB$~\cite{Schaefer:2007pw}.

Our idea proposed here is to make use of Fig.~\ref{fig:entropy} to fix
$T_0(\muB)$ for consistency with phenomenology.  [One can pick up
  other thermodynamic quantities than the entropy density like the
  internal energy density, which would anyway make little change in
  the final result.]  In Ref.~\cite{Cleymans:2005xv} the freeze-out
curve is parametrized as $T_{\rm f}(\muB)=a-b\muB^2-c\muB^4$ with the
fitting result $a=166(2)\MeV$, $b=1.39(16)\times10^{-4}\MeV^{-1}$, and
$c=5.3(21)\times10^{-11}\MeV^{-3}$.  Because the behavior of the
entropy density must be dominantly controlled by deconfinement, we
postulate that the change in $T_0$ is to be correlated with
$T_{\rm f}(\muB)$.  [We see that the freeze-out points and the entropy
  band in Fig.~\ref{fig:freezeout} have roughly same curvature.]  Let
us simply use same $b$ and make an ansatz as
\begin{equation}
 \frac{T_0(\muB)}{T_0} = 1 - (b T_0) \Bigl(\frac{\muB}{T_0}\Bigr)^2
  \;,
 \label{eq:T0}
\end{equation}
which yields the blue band in the middle of Fig.~\ref{fig:entropy}.
[To prevent unphysical negative $T_0$ for large $\muB$ we set a
  threshold at $10\MeV$ so that $T_0 \geq 10\MeV$.  Hence, the results
  at $T<10\MeV$ are not meaningful.]  We see at a glance that the
results from this modified PNJL model have a reasonable overlap with
the Statistical Model results in the whole density region as plotted.

At this point one might have thought that the energy scale in the
quark (NJL) sector should be modified as well.  We will come back to
this question after discussing the phase diagram next.
\vspace{3mm}


\paragraph*{Phase Diagram}

Now we get ready to proceed to the possible QCD phase diagram that is
at least consistent with the Statistical Model outputs in
Fig.~\ref{fig:freezeout}.  Using the standard computational procedure
of the mean-field PNJL model we can solve
$\langle\bar{\psi}\psi\rangle$ and $\Phi$ as functions of $T$ and
$\muB$, from which the phase boundaries of chiral restoration and
deconfinement are located.


\begin{figure}
 \includegraphics[width=\columnwidth]{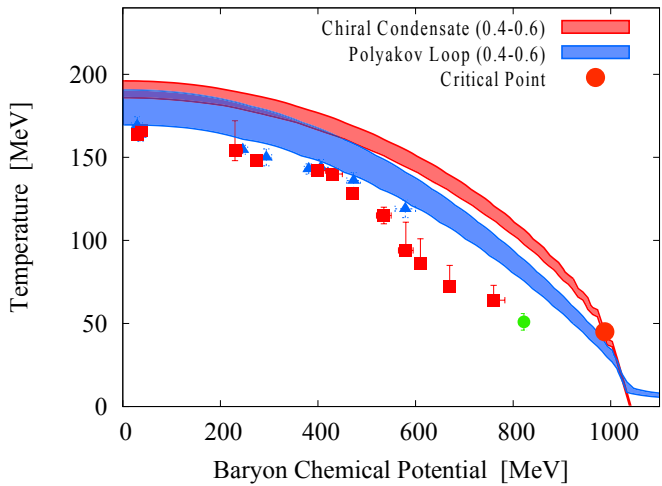}
 \caption{Phase boundaries associated with deconfinement (blue band)
   and chiral restoration (red band).  Each band represents a region
   where the (normalized) order parameter develops from $0.4$ to
   $0.6$.}
 \label{fig:diagram}
\end{figure}


Figure~\ref{fig:diagram} shows the phase diagram from the modified
PNJL model.  The blue (red) band is a region where the Polyakov loop
(normalized light-quark chiral condensate) increases from $0.4$ to
$0.6$.  In contrast to the old PNJL model ones, the new results show
that the chiral phase transition is almost parallel to and entirely
above the deconfinement, which agrees with the situation considered
recently in Ref.~\cite{Castorina:2010gy}.  We have found the critical
point at $(\muB,T)\simeq(45\MeV,330\MeV)$, but would not take it
seriously since its location is easily
affected~\cite{Fukushima:2003fw}.  Still, it is a good news for the
critical point search that two QCD phase transitions stay close to
each other, for the experimental signature would be detectable only if
the critical point sits sufficiently near the freeze-out point.
\vspace{3mm}


\paragraph*{Discussions}

It is an intriguing observation that the chiral phase transition
occurs later than deconfinement.  This is quite consistent with the
Statistical Model assumption.  In the Statistical Model the hadron
masses are just the vacuum values and any hadron mass/width
modifications are neglected, which would be a reasonable treatment
only if the chiral phase transition is above the Hagedorn
temperature.  Under such a phase structure, besides, our assumption of
neglecting $\muB$-dependence in the NJL-model parameters turns out to
be as acceptable as the Statistical Model treatment.  This can be
understood from the fact that the NJL part yields the hadron masses in
the vacuum which are intact in the Statistical Model.

The failure of the standard PNJL model is attributed to baryonic
degrees of freedom missing in a non-confining quark description.
Hence, one may say that a modification made in $\mathcal{U}[\Phi]$
stems from such crossover between baryons and quarks, which is
presumably parametrized by the Polyakov loop alone, similarly to the
transverse gluon pressure.  It is an important question how our
phenomenological input~\eqref{eq:T0} is validated/invalidated from the
first-principle QCD calculation, which will be answered by future
developments in the functional renormalization group
method~\cite{Braun:2007bx}.

Finally, our conclusion is that, if quarkyonic matter is defined by
restored chiral symmetry with confinement, it is not consistent with
the Statistical Model and is unlikely to occur.  However, to complete
our analysis it should be necessary to think of the quarkyonic
spiral~\cite{Kojo:2009ha}, which is an important future problem.
\vspace{3mm}


\paragraph*{Acknowledgments}

The author thanks Y.~Hidaka for numerical assistance.  He also thanks
K.~Redlich and J.M.~Pawlowski for discussions and A.~Andronic for the
numerical data of his freeze-out points.  This work is supported by
Japanese MEXT grant No.\ 20740134 and in part by Yukawa International
Program for Quark Hadron Sciences.


\end{document}